\documentclass{article}

\usepackage{microtype}
\usepackage{graphicx}
\usepackage{subfigure}
\usepackage{booktabs} % for professional tables

\usepackage{hyperref}

\makeatletter
\newcommand\footnoteref[1]{\protected@xdef\@thefnmark{\ref{#1}}\@footnotemark}
\makeatother

\usepackage[accepted]{icml2019}

\icmltitlerunning{Learning to Groove}

\begin{document}

\twocolumn[
\icmltitle{Learning to Groove with Inverse Sequence Transformations}



\icmlsetsymbol{equal}{*}

\begin{icmlauthorlist}
\icmlauthor{Jon Gillick}{ischool,goog}
\icmlauthor{Adam Roberts}{goog}
\icmlauthor{Jesse Engel}{goog}
\icmlauthor{Douglas Eck}{goog}
\icmlauthor{David Bamman}{ischool}
\end{icmlauthorlist}

\icmlaffiliation{ischool}{School of Information, University of California, Berkeley, CA, U.S.A}
\icmlaffiliation{goog}{Google AI, Mountain View, CA, U.S.A}

\icmlcorrespondingauthor{Jon Gillick}{jongillick@berkeley.edu}

\icmlkeywords{Machine Learning, ICML, Music, Drums, Seq2Seq, Variational Information Bottleneck}

\vskip 0.3in
]



\printAffiliationsAndNotice  % leave blank if no need to mention equal contribution

\begin{abstract}

We explore models for translating abstract musical ideas (scores, rhythms) into expressive performances using Seq2Seq and recurrent variational Information Bottleneck (VIB) models. Though Seq2Seq models usually require painstakingly aligned corpora, we show that it is possible to adapt an approach from the Generative Adversarial Network (GAN) literature (e.g., Pix2Pix \cite{isola2017image} and Vid2Vid \cite{wang2018video}) to sequences, creating large volumes of paired data by performing simple transformations and training generative models to plausibly invert these transformations. Music, and drumming in particular, provides a strong test case for this approach because many common transformations (quantization, removing voices) have clear semantics, and models for learning to invert them have real-world applications.  Focusing on the case of drum set players, we create and release a new dataset for this purpose, containing over 13 hours of recordings by professional drummers aligned with fine-grained timing and dynamics information.  We also explore some of the creative potential of these models, including demonstrating improvements on state-of-the-art methods for Humanization (instantiating a performance from a musical score).

\end{abstract}

\section{Introduction}

%Artists, musicians, writers, and creators of all kinds describe their day-to-day work as a process of translating ideas into finished pieces~\cite{}.  Creators of digital \textbf{tools} to support this work seek to make the translation process smoother by allowing access... Researchers in HCI, Intelligent User Interfaces, Computer Graphics, Natural Language Processing, Music Information Retrieval, and many other fields are concerned with building and evaluating digital tools to support these processes.  In this work, we are interested in studying tasks that may fit into these practical toolkits in the near future.  We focus our investigation on the task of generating drum performances from scores because while it has been shown to be of practical interest to musicians and music software developers (as evidenced by the existence of tools for the task already being embedded in industry standard music production software), the task has received limited attention within the machine learning community.

A performance can be viewed as a translation of an idea conceived in the mind to a finished piece on the stage, the screen, or the speakers.  The long-standing goal of many creative technologies is to enable users to render realistic, compelling content that brings an idea to life; in so doing, finding a balance between realism and control is important.  This balance has proved difficult to achieve when working with deep generative models, motivating recent work on conditional generation in several modalities including images~\cite{yan2016attribute2image}, speech~\cite{shen2018natural}, and music~\cite{simon2018learning}.  In this work, rather than generating new content conditioned on one of a fixed set of classes like \emph{rock} or \emph{jazz}, we are interested in learning to translate ideas from representations that are more easily expressed (musical abstractions such as scores) into instantiations of those ideas that would otherwise be producible only by those skilled in a particular instrument (performances).

We use the metaphor of translation from idea to finished work as a starting point for our modeling choices, adapting and modifying Seq2Seq models typically used in machine translation \cite{sutskever2014sequence}.  
While musical scores and performances can be thought of as different expressions of the same idea, our setting differs from translation in that
%, akin to sentences in different languages
musical scores are designed to be compressed representations; the additional information needed to create a performance comes from the musician.  In this work, we set up a data collection environment in which a score can be deterministically extracted from the performance in a manner consistent with the conventions of western music notation, effectively yielding a parallel corpus.  Furthermore, though western music notation is well established as one compressed representation for music, our data allows us to explore other representations that are compressed in different ways; we propose and explore two such transformations in this work, which we call Infilling and Tap2Drum.  Learning to map from these reduced versions of musical sequences to richer ones holds the potential for creative application in both professional and amateur music composition, production, and performance environments.

\begin{figure*}[t!]
    \begin{centering}
    \includegraphics[scale=0.25]{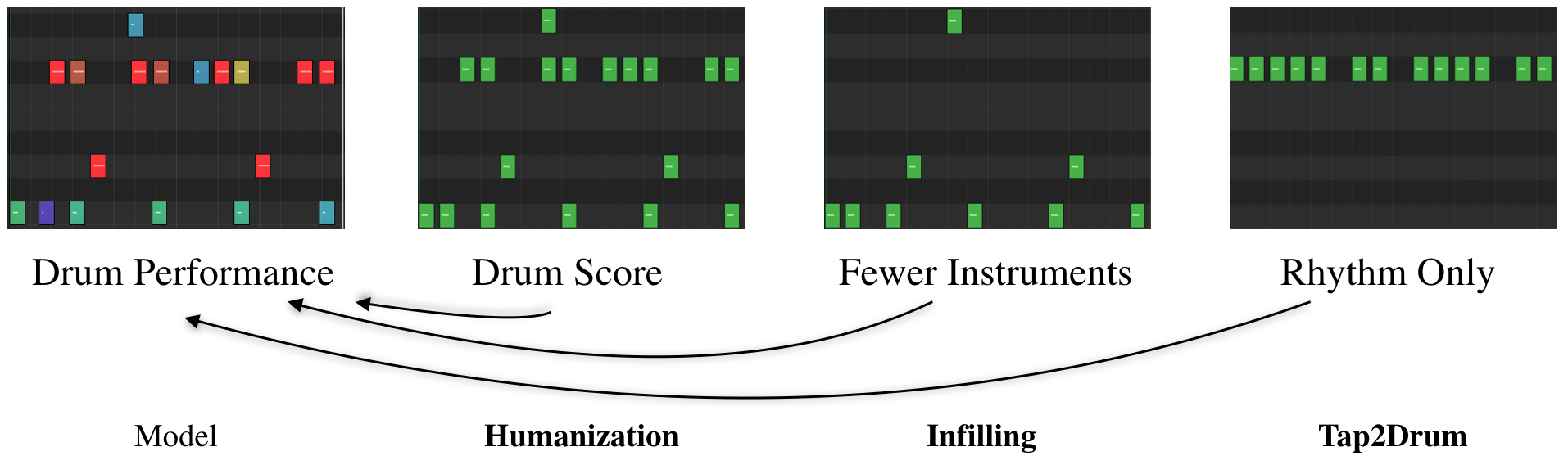}
    \caption{Learning inverse sequence transformations for drumming. Moving from left to right, the representations become progressively simpler, first removing expressive timing (small shifts off the grid) and dynamics (color, with higher velocities in red), then removing one of the voices, and then compressing all voices to a single track. We train models to map from each of these deterministically compressed representations back to complete realizations of drum performances. The inverse transformations correspond to Humanization, Infilling, and Tap2Drum respectively, and require progressively easier inputs for an untrained user to create. }
    \label{fig:model_inputs}
    \end{centering}
\end{figure*}

We focus in this work specifically on drums; though drums and percussion are essential elements in many styles of modern music, creating expressive, realistic sounding digital drum performances is challenging and time consuming.  Humanization functions have been embedded in industry standard music production software for many years, but despite evidence that the current methods used in professional toolkits (randomly jittering note timings and dynamics with Gaussian noise) have little effect on listener preferences~\cite{senn2018groove}, machine learning based methods have not yet made their way into many mainstream environments to replace them.  We hope that our data, models, and methods for generating and controlling drum performances will continue to drive forward the growing body of work on expressive performance modeling.

In this work we make the following contributions:
\begin{itemize}
    \item We collect a new dataset an order of magnitude larger than the largest previously publicly available, with 13.6 hours of recordings of 10 drummers playing electronic drum kits instrumented with sensors to capture precise performance characteristics in MIDI format. We pair this data with associated metadata including anonymized drummer identifiers, musical style annotations, and tempo, while also capturing and aligning the synthesized audio outputs.
    
    \item We present a data representation and a class of models that we call GrooVAE.  We use our models to explore the task of Humanization, learning to perform a musical score for drum set, demonstrating improvements over previous methods.
    
    \item We introduce, implement, and evaluate two new tasks made possible by our data and model, which we call Drum Infilling and Tap2drum.  We argue that these models, along with Humanization, may allow for user control over realistic drum performance generation without expertise playing the drum set.
\end{itemize}

Code, data, trained models, and audio examples are available at \url{https://g.co/magenta/groovae}.
%\footnote{Online Resources: \newline 
%Code: \url{http://goo.gl/magenta/musicvae-py} \newline
%Notebook: \url{http://goo.gl/magenta/groovae-colab} \newline
%Data: \url{https://magenta.tensorflow.org/datasets/groove} \newline
%Audio: \url{https://goo.gl/magenta/groovae-examples} \newline
%}

\section{Related Work}

A small number of previous studies explore machine learning methods for generating expressive drum performance timing, employing linear regression and K-Nearest Neighbors~\cite{wright2006towards}, or Echo State Networks~\cite{tidemann2009groovy}.  These studies use data from different musical genres and different drummers, so relative performance between methods is not always clear.  In most cases, however, listening tests suggest that qualitative results are promising and can produce better outputs than those created heuristically through a \emph{groove template}\footnote{Groove templates, which are used commonly in music production practice, copy exact timings and velocities of notes from a template sequence.}~\cite{wright2006towards}.

Other work on expressive performance modeling focuses on piano rather than drums, leveraging data from performances recorded on electronic keyboards or  Disklaviers, pianos instrumented with MIDI inputs and outputs~\cite{gu2012modeling, gu2013creating, oore2018time, huang2018music, hawthorne2018enabling}.  Recent impressive results in generating both MIDI and audio also suggest that given enough data, neural sequence models can realistically generate expressive music.  One drawback of the large piano datasets, however, is that they lack gold standard alignments with corresponding musical scores, making tasks like Humanization more challenging.

There are of course many other settings besides music in which learning to translate from abstractions to instantiations can be useful.  State-of-the-art methods for speech synthesis~\cite{wang2018style, stanton2018predicting}, and story generation~\cite{fan2018hierarchical} typically use Seq2Seq frameworks.  Unlike our case, however, these methods do require paired data, though some recent work attempts to reduce the amount of paired data needed through self-supervised learning~\cite{chung2018semi}.

Perhaps most similar to our setting is recent work in the image domain, which has demonstrated the ability of GAN models to translate from simple, potentially user-provided, inputs into photo-realistic outputs~\cite{isola2017image, wang2018video}.  Images and music are similar in that their contents can survive abstraction into simplified versions through lossy transformations like quantization or edge detection while still retaining important semantic details.
%, as opposed to a sentence, which is more likely to lose its meaning if words are removed.  
Images, however, are structured fundamentally differently than musical sequences and tend to benefit from different modeling choices -- in particular the use of GANs, which have not been demonstrated to work as well for music as recurrent neural networks.

\section{Data}

Existing work on expressive drum modeling focuses only on small datasets with a limited number of sequences, drummers and genres~\cite{wright2006towards, tidemann2008groovy}.  Other studies that model expressive performance on different instruments (typically piano) use larger and more diverse datasets~\cite{huang2018music, simon2017performance, hawthorne2018enabling}, but these data lack ground truth alignments between scores and performances; this alignment, which allows use to measure time relative to a metronome, is key to the applications we explore in this work.  Several companies also sell drum loops played to a metronome by professional drummers, but these commercially produced loops may be edited in post-production to remove human error and variation, and they also contain restrictive licensing agreements that prohibit researchers from sharing their models.

There is currently no available gold standard dataset that is of sufficient size to reasonably train modern neural models and that also contains a precise mapping between notes on a score and notes played by a performer.  

\subsection{Groove MIDI Dataset}

To enable new experiments and to encourage comparisons between methods on the same data, we collect a new dataset of drum performances recorded in MIDI format (the industry standard format for symbolic music data) on a Roland TD-11\footnote{\url{https://www.roland.com/us/products/td-11/}} electronic drum kit.  MIDI notes (we also refer to them as hits) are each associated with an instrument, a time, and a velocity.  Microtimings, (we also call them timing offsets), describe how note timings stray from a fixed grid, and velocities (or dynamics) denote how hard notes are struck.  Taken together, we refer to microtiming and velocity as \emph{performance characteristics} or \emph{groove}, and the quantized times of the notes define a musical score (also called a pattern or sequence).   While some  nonpercussive instruments like strings or horns, which allow for continuous changes to a single note, are difficult to represent with MIDI, many styles of drum set playing can be well specified through microtiming and velocity.

The dataset, which we refer to as the Groove MIDI Dataset (GMD), is publicly available for download at \url{https://magenta.tensorflow.org/datasets/groove}. The GMD contains 13.6 hours, 1,150 MIDI files, and over 22,000 measures of tempo-aligned expressive drumming, making it an order of magnitude larger than the largest comparable dataset. Complete details of acquisition and annotation can be found in  Appendix A.

\subsection{Preprocessing} \label{ref:ssec:preprocessing}

Though the Groove Midi Dataset contains all the information captured by the electronic drum kit, including multiple sensors to detect hits on different parts of each drum, we make several preprocessing choices to simplify our models for this work.  First, we map all drum hits to a smaller set of 9 canonical drum categories, following~\citet{roberts2018hierarchical}.  These categories represent the most common instruments in standard drum kits: bass drum, snare drum, hi-hats, toms, and cymbals; we display the full list of drum categories along with their respective MIDI mappings in Appendix B.

After partitioning recorded sequences into training, development, and test sets, we slide fixed size windows across all full sequences to create drum patterns of fixed length; though we explored models for sequences of up to 16 measures, for consistency we use 2 measure (or 2 \emph{bar}) patterns for all reported experimental evaluations, sliding the window with a hop size of 1 measure.  We chose 2 measures for our experiments both because 2 bars is a typical length for drum loops used in music production practice and because these sequences are long enough to contain sufficient variation but short enough to quickly evaluate in listening tests.

As a final step, motivated by the fact that music production software interfaces typically operate at 16th note resolution~\cite{wherry2006quantise}, we take 16th notes as the fundamental timestep of our data. Each drum hit is associated with the closest 16th note metrical position; if multiple hits on the same drum category map to the same timestep, we keep the loudest one.  Although this preprocessing step forces us to discard some of the subtle details of drum rolls that can be played on a single drum faster than 16th notes, we found that perceptually, much of the expressiveness in drumming can be conveyed at this resolution.  Moreover, after experimenting both with finer resolutions (32nd or 64th notes) and data representations that count time in absolute time (milliseconds) rather than relative time (as in~\citet{simon2017performance}), we found that the gains in modeling yielded by this constraint were more important than the details lost.  One potential path forward in future work might be to supplement our data representation with an explicit token for a drum roll.

\subsection{Data Represention}

After preprocessing, our data points are of fixed length: each sequence has $T$ timesteps (one per 16th note) and $M$ instruments per timestep.  The full representation consists of the below three $T\times M$ matrices, with values $T=32$ and $M=9$ for all reported experiments.

\paragraph{Hits.}  To represent the presence or absence of drum onsets, or hits, in a sequence, we define a binary-valued matrix ${H}$, which contains all the information in a basic drum score.  A column of ${H}$ contains the drum score for one of the nine instruments in the drum set, and a row of ${H}$ contains the drum score for all nine instruments at a single timestep.
    
\paragraph{Offsets.} A continuous-valued matrix ${O}$ stores the timing offsets, taking values in [-0.5, 0.5) that indicate how far and in which direction each note's timing lies relative to the nearest 16th note.  Drum hits may fall at most halfway between their notated position in time and an adjacent position.  We can examine ${O}$ to compute statistics on microtiming: positive values indicate playing behind the beat (late); negative values demonstrate playing ahead (early).  

\begin{figure}
    \begin{centering}
    \includegraphics[scale=0.22]{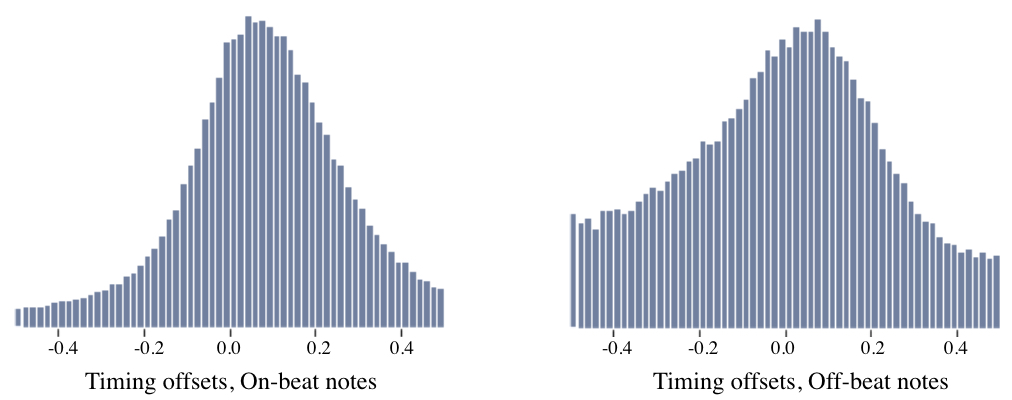}
    \caption{Distribution of timing offsets for notes in the training set.  On-beat notes (landing on an eighth note), shown on the left, are more often played late, whereas off-beat notes (not landing on an eighth note), on the right, are more often played early.}
    \label{fig:timing_offsets}
    \end{centering}
\end{figure}

Modeling continuous as opposed to discrete values for offsets allows us to take advantage of the fact that timing appears to be approximately normally distributed at any given metrical position (as shown in Figure~\ref{fig:timing_offsets}); intuitively, models should be penalized more for predictions that are further from the ground truth.  We experimented with various continuous and discrete representations including logistic mixtures~\cite{salimans2017pixelcnn++}, thermometer encodings~\cite{buckman2018thermometer}, and label smoothing~\cite{pereyra2017regularizing}, but we found that modeling timing offsets and velocity as single Gaussian distributions (conditional on the LSTM state) produced by far the most perceptually realistic results.
    
\paragraph{Velocities.} Another continuous-valued matrix ${V}$ stores the velocity information (how hard drums are struck).  We convert velocity values from the MIDI domain (integers from 0-127) to real numbers in [0,1].

%\subsection{Drummer Perspectives}

%While much work in generative models relies on existing data that has been re-purposed for research, bringing in professional drummers to record provided us with an opportunity to ask the actual creators of the content about their perspectives on the process of contributing to such a dataset and on potential applications.  We conducted short semi-structured interviews with four hired drummers to gain insight into their perspectives on this collaboration between artists and machine learning research. 

%Though the focus of this paper is on the modeling, we briefly report one common theme that surfaced during interviews: t

%\begin{itemize}
%    \item 3 of 4 hired drummers expressed having mixed feelings about contributing to the project
%    \item 2 explicitly mentioned that their attitudes toward the technology changed over the course of the recording sessions
%    \item All 4 said they enjoyed being hired for the gig, and 2 said they ished they could do it more often
%    \item include quotes
%\end{itemize}

\section{Modeling Objectives}

%Though the models we describe may be of practical use for other other related tasks, 
We focus our experiments and analysis on three particular applications of expressive performance modeling.  For audio examples of additional tasks such as unconditional sampling, interpolation, and style transfer, see the online supplement\footnote{\label{fn:supp}\url{http://goo.gl/magenta/groovae-examples}}.

\paragraph{Humanization.} Our first objective is to generate, given a 16th-note-quantized drum pattern with no microtiming or velocity information (i.e., a drum score), a MIDI performance of that score that mimics how a professional drummer might play it.  Because this task has an existing body of work, we focus most of our experiments and evaluations on this task.%Since there are many more possible performances than there are scores, stochastic outputs are acceptable and, for practical purposes, indeed desirable.  %Heuristic methods for Humanization, which usually work by randomly jittering the timings and velocities of notes with Gaussian noise, have long been available within industry standard music production software such as Logic Pro and Ableton Live.  

\paragraph{Infilling.} We introduce a second task of interest within the same contexts as Humanization that we call Drum Infilling. The objective here is to complete or modify a drum beat by generating or replacing the part for a desired instrument.  We define an instrument as any one of the 9 categories of drums and train models that learn to add this instrument to a performance that lacks it.  For brevity, we choose a single drum category (hi-hat) as a basis for our evaluations.  Infilling provides a case for examining computer assisted composition, allowing a composer to sketch parts for some pieces of the drum kit and then receive suggestions for the remaining parts.  Previous work explores Infilling in the context of 4-part Bach compositions ~\cite{huang2019counterpoint} and in piano performance ~\cite{ippolitoinfilling}; we look at the task for the first time in the context of drums.
%similar to text generation conditioned on a prompt~\cite{serban2016building, wen2015semantically}, infilling allows a composer to sketch parts for some pieces of the drum kit and then receive suggestions for the remaining parts.

\paragraph{Tap2Drum.} In this last task, we explore our models' ability to generate a performance given an even further compressed musical representation.  While western musical scores usually denote the exact notes to be played but lack precise timing specifications, we propose a new representation that captures precise timing but does not specify exactly which notes to play.  In this setting, which we call Tap2Drum, we give our model note offset information indicating the microtiming, but we do not specify the drum categories or velocities as inputs, leaving the decision of which instrument to hit and how hard to the model.  Because almost anyone can tap a rhythm regardless of their level of musical background or training, this input modality may be more accessible than musical notation
%potentially lowering the bar of entry 
for those who would like to express their own musical ideas on a drum set but lack the skills of a drummer.

%While the practical utility of Infilling and Tap2Drum remain to be determined, we include them in our discussion as examples of potentially novel applications of this line of research.

\section{Models}

We compare several models for Humanization, selecting the best performing one for our experiments with Infilling and Tap2Drum.  

\subsection{Baselines}

For our baseline models, we focus on models from the literature that have been used before for Humanization in the context of drum performances aligned to a metronome.

\subsubsection{Quantized}
As a simple baseline, we set all offsets to 0 and velocities to the mean value in the training set.

\subsubsection{Linear Regression}
For this baseline, we regress $H$ against $V$ and $O$, predicting each element of $V$ and $O$ as a linear combination of the inputs $H$.

\subsubsection{K-Nearest Neighbors} \label{ssec:knn}
\citet{wright2006towards} report strong results in using K-Nearest Neighbors to predict microtiming in Brazilian percussion.  They define a hand-crafted distance measurement between notes, retrieve the $K$ notes in the training set nearest to a given note in a test sequence, and then take the mean timing offset of those notes.  Their definition of nearest notes, however, requires that the same sequence appear in both training and test sets.  Since our test set emphasizes unseen sequences, we adapt the method as follows: first we retrieve the $K$ nearest \emph{sequences}, measuring distance $D_{i,j}$ by counting the number of notes in common between a test sequence ${x_i}$ and each training sequence ${x_j}$, which can be computed easily through the Hadamard product of their respective binary matrices, ${H_i}$ and ${H_j}$: 
\begin{equation}
D_{i,j} = \sum H_i \circ H_j  
\end{equation}
Given the closest $K$ sequences $[S_1, \ldots, S_K]$, we then compute predicted velocities ${\hat{V}}$ and offsets ${\hat{O}}$ by taking the element-wise means of the corresponding ${V}$ and ${O}$ matrices:
\begin{equation}
\hat{V} = \frac{1}{K}\sum\limits_k V_k  
\end{equation}
\begin{equation}
\hat{O} = \frac{1}{K}\sum\limits_k O_k  
\end{equation}
When reconstructing a MIDI sequence, we ignore the entries of $\hat{V}$ and $\hat{O}$ for which the corresponding entry of $H$ is 0.

Choosing ${K=1}$ is equivalent to selecting the most similar sequence as a groove template, and choosing $K$ to be the cardinality of the training set yields a single groove template that summarizes the average performance characteristics of the entire set.  Through a grid search on the development set, we found that setting ${K=20}$ performed best, close to the reported ${K=26}$ from~\citet{wright2006towards}.

\begin{figure}
    \begin{centering}
    \includegraphics[scale=0.16]{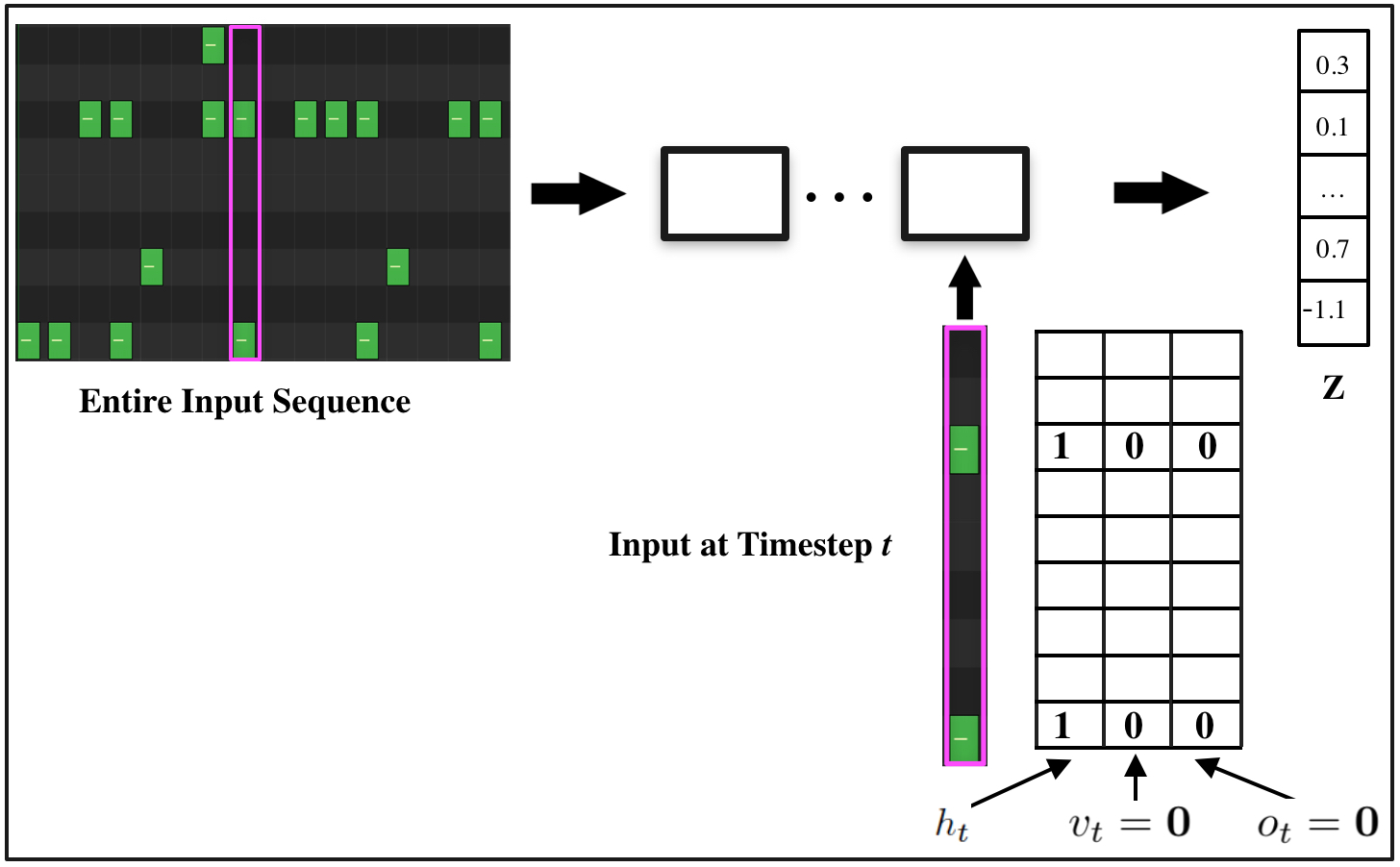}
    \caption{The forward direction of our encoder architecture for the Seq2Seq Humanization model.  Input sequences are visualized as piano rolls, with drum categories on the vertical axis and time on the horizontal axis. LSTM inputs are shown for a single timestep $t$.  Instruments with no drum hits at time $t$ are shown as blank, although for implementation we fill these blank cells with 0's.  Note that no velocity or timing offset information is passed to the encoder.}
    \label{fig:encoder}
    \end{centering}
\end{figure}

\subsection{Proposed Models}

\subsubsection{MLP}

To train multilayer perceptron (MLP) neural networks for Humanization, we concatenate the matrices $H$, ${V}$, and ${O}$ to form a target matrix $y \in {R}^{T\times(M*3)}$.  We pass $H$ into the model as inputs, training the network to minimize the squared error between $y$ and predictions $\hat{y}$.  For the MLP, we use a single hidden layer of size 256 and ReLU nonlinearities.  We train all our neural models with Tensorflow~\cite{abadi2016tensorflow} and the Adam optimizer~\cite{kingma2014adam}. 

\subsubsection{Seq2Seq} \label{ssec:Seq2Seq}

Sequence to sequence models~\cite{sutskever2014sequence} encode inputs into a single latent vector, typically with a recurrent neural network, before autoregressively decoding into the output space.  For this architecture, we process the drum patterns over $T=32$ timesteps, encoding a drum score to a vector $z$ with a bidirectional LSTM and decoding into a performance with a 2-layer LSTM.

\paragraph{Encoder} The encoder is based on the bidirectional LSTM architecture used in~\citet{roberts2018hierarchical}, though 
%because we have less data,
we change the LSTM layer dimensions from 2048 to 512 and the dimension of $z$ from 512 to 256.  At each timestep $t$, we pass a vector $h_t$, which is row $t$ of $H$, to the encoder, representing which drums were hit at that timestep; velocities and timing offsets are not passed in.  As shown in Figure~\ref{fig:model_inputs}, we keep the same architecture for Infilling and Tap2Drum, only modifying the inputs to switch tasks.  Figure~\ref{fig:encoder} demonstrates one step of the forward direction of the encoder.

\paragraph{Decoder} We use a 2-layer LSTM of dimension 256 for our decoder, which we train to jointly model $H$, $V$, and $O$.  Unlike~\citet{roberts2018hierarchical}, however, we split the decoder outputs at each timestep $t$ into 3 components, applying a softmax nonlinearity to the first component to obtain a vector of predicted hits $\hat{h_t}$, sigmoid to the second component to get velocities $\hat{v_t}$, and tanh to the third, yielding timing offsets $\hat{o_t}$.  These vectors are compared respectively with ${h_t}$, ${v_t}$, and ${o_t}$, the corresponding rows of ${H}$, ${V}$, and ${O}$, and finally summed to compute the primary loss for this timestep $L_t$:

\begin{equation}
L_t = CrossEntropy(h_t, \hat{h_t}) + (v_t - \hat{v_t})^2 + (o_t - \hat{o_t})^2
\end{equation}

We train the model end to end with teacher forcing.

\begin{figure}
    \begin{centering}
    \includegraphics[scale=0.15]{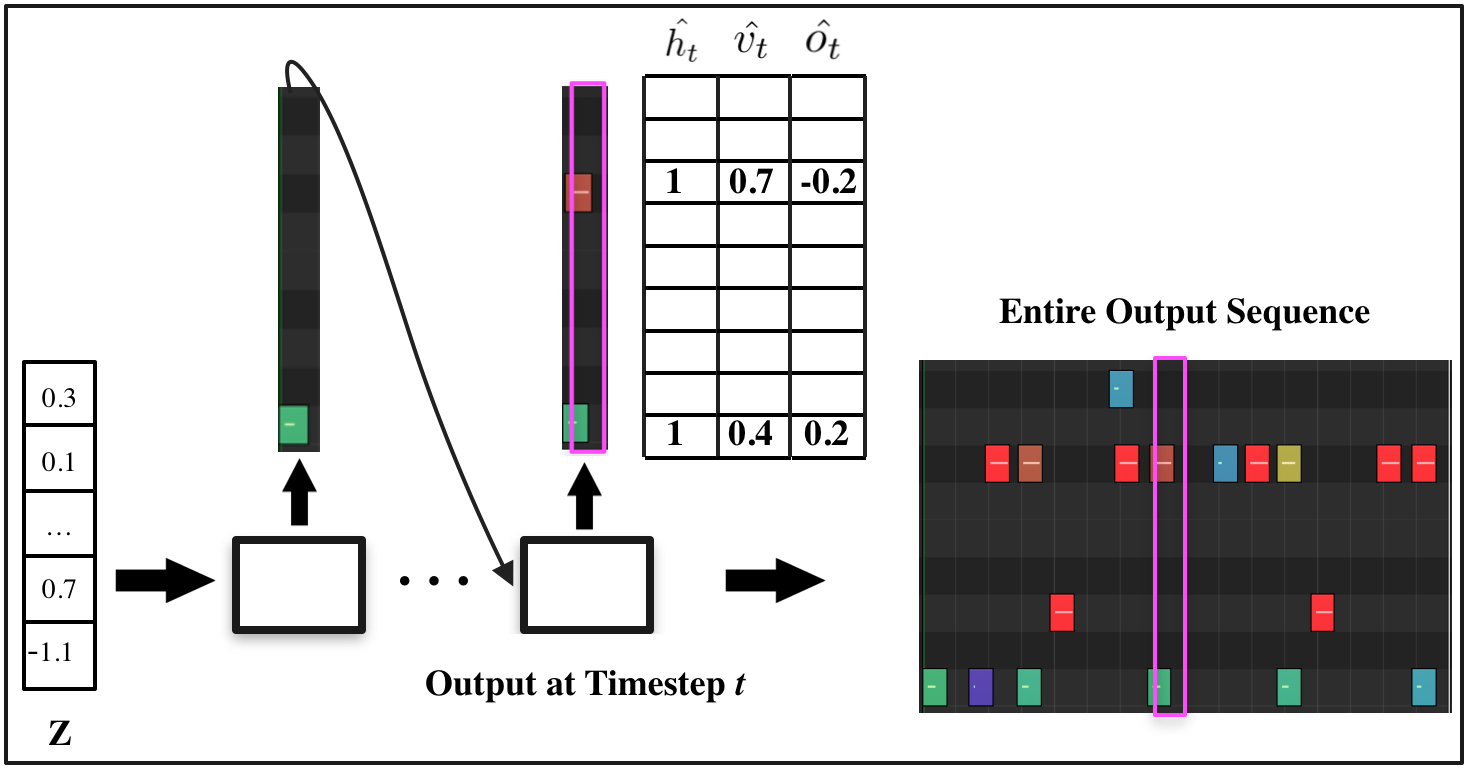}
    \caption{Decoder architecture for Seq2Seq Humanization model.  The decoder generates outputs for drum hits, velocities and timing offsets. Velocity is visualized in color, and output notes appear slightly earlier than the grid lines, indicating negative offsets.}
    \label{fig:decoder}
    \end{centering}
\end{figure}

%\subsubsection{Variational Information Bottleneck}
%\jgcomment{@adam, @jesse - I could use help making sure this section makes sense and has the right citations.}
%Because we are interested in creative applications of our models, we would like our encoded vectors to be distributed in such a way as to allow for operations like vector arithmetic and interpolation.  Adding a variational loss~\cite{kingma2013auto} to encourage the embeddings $z$ to lie close to a prior (in our case multivariate normal) distribution, has been shown both to be a strong regularizer and an effective way of enabling these vector operations.

%In our Seq2Seq models, we add a variational loss using the same hyperparameters as~\citet{roberts2018hierarchical}.  Though we focus in this work on studying the particular tasks of Humanization, Infilling, and Tap2drum rather than on the the geometry and optimization of latent spaces, we include audio examples in the supplementary material demonstrating the ability of our models to generate new samples and interpolate between them.  Note that if we include the performance characteristics in the model inputs by $v$ and $o$ into the encoder along with $h$, this model becomes a Variational Autoencoder.

\subsubsection{Groove Transfer}

%The decoder described in Section \ref{ssec:Seq2Seq} learns to generate ${H}$, ${V}$, and ${O}$ starting with only $z$ as its initial input.  Since the task at hand, however, is to generate ${V}$ and ${O}$ given $H$, a decoder should not necessarily be required to reconstruct $H$.  Working from this intuition, 
We experiment with one more model that we call Groove Transfer.  This architecture is identical to our Seq2Seq model except that at each timestep $t$ we concatenate $h_t$, the vector for the hits at time $t$, to the decoder LSTM inputs using the conditioning procedure of~\citet{simon2018learning}.  By allowing the decoder to learn to copy $h_t$ directly to its outputs, we incentivize this encoder to ignore $H$ and only learn a useful representation for generating ${V}$ and ${O}$.  The main benefit of this architecture over Seq2Seq is that the modification allows us to disentangle the performance characteristics (the groove) of a sequence $S_1$ from the score $H_1$, capturing the performance details in the groove embedding $z_1$.  We can then pass $z_1$ to the decoder along with the content $H_2$ of another sequence $S_2$ to do style transfer for drum performances.  Audio examples of Groove Transfer can be found in the supplementary materials\footnoteref{fn:supp}.

%Under this setup, we found that the the decoder ignores the conditioning inputs from $H_2$ and instead accesses $H_1$ via $z_1$. We therefore add a variational loss term to this model, turning our Seq2Seq model into a Variational Information Bottleneck~\cite{alemi2016deep} and training the embeddings $z$ to lie close to a prior (multivariate normal) distribution.  This modification allows us to disentangle the performance characteristics (the groove) of a sequence $S_1$ from the score $H_1$, capturing the performance details in the groove embedding $z_1$.  We can then pass $z_1$ to the decoder along with the content $H_2$ of another sequence $S_2$.  Following~\citet{roberts2018hierarchical}, we train by maximizing a modified Evidence Lower Bound (ELBO) using the hyperparameter $\beta = 0.2$. 

We also apply Groove Transfer to Humanization as follows: given a score $H_2$, we embed the closest $k=3$ sequences in the training set as defined by the distance metric in Section \ref{ssec:knn}, store the mean of the $k$ embeddings in a vector $z_k$, and then transfer the groove vector $z_k$ to $H_2$.

%Figure~\ref{fig:groove_transfer} demonstrates the architecture of the Groove Transfer decoder.  

\subsection{Variational Information Bottleneck}

Our test data, while disjoint from the training data, comes from the same set of drummers, and its distribution is meant to be similar.  In the real world, however, we would like to be able to trade off between realism and control; when faced with a very unlikely drum sequence, such as one quickly sketched in a music production software interface, we may want to choose a model that constrains its output to be close to the realistic examples in the training set, potentially at the cost of changing some of the input notes.  To this end, we add a variational loss term to both Seq2Seq and Groove Transfer, turning the models into a Variational Information Bottleneck (VIB) ~\cite{alemi2016deep} and training the embeddings $z$ to lie close to a prior (multivariate normal) distribution. Following~\citet{roberts2018hierarchical}, we train by maximizing a modified Evidence Lower Bound (ELBO) using the hyperparameter $\beta = 0.2$.  We report our quantitative metrics both with and without the VIB.

%\begin{figure}
%    \begin{centering}
%    \includegraphics[scale=0.17]{images/_groove_transfer.jpg}
%    \caption{Decoder architecture for Groove Transfer model. $z$ is %generated by the encoder as in normal Humanization, but the hit matrix $H$ %is also given to the decoder.}
%    \label{fig:groove_transfer}
%    \end{centering}
%\end{figure}
 
\section{Results}

%Since Humanization is an established task with existing baselines for comparison, we focus our evaluations here, subsequently applying the best performing models to Infilling and Tap2Drum.

\subsection{Listening Tests}

As is the case with many generative models, especially those designed for creative applications, we are most interested in the perceptual quality of model outputs; for this reason, we also highly encourage the reader to listen to the audio examples in the supplementary materials\footnoteref{fn:supp}.  In our setting, high quality model outputs should sound like real drum performances.  We examine our models through multiple head-to-head listening tests conducted on the Amazon Mechanical Turk platform.

\begin{figure}
    \begin{centering}
    \includegraphics[width=\columnwidth]{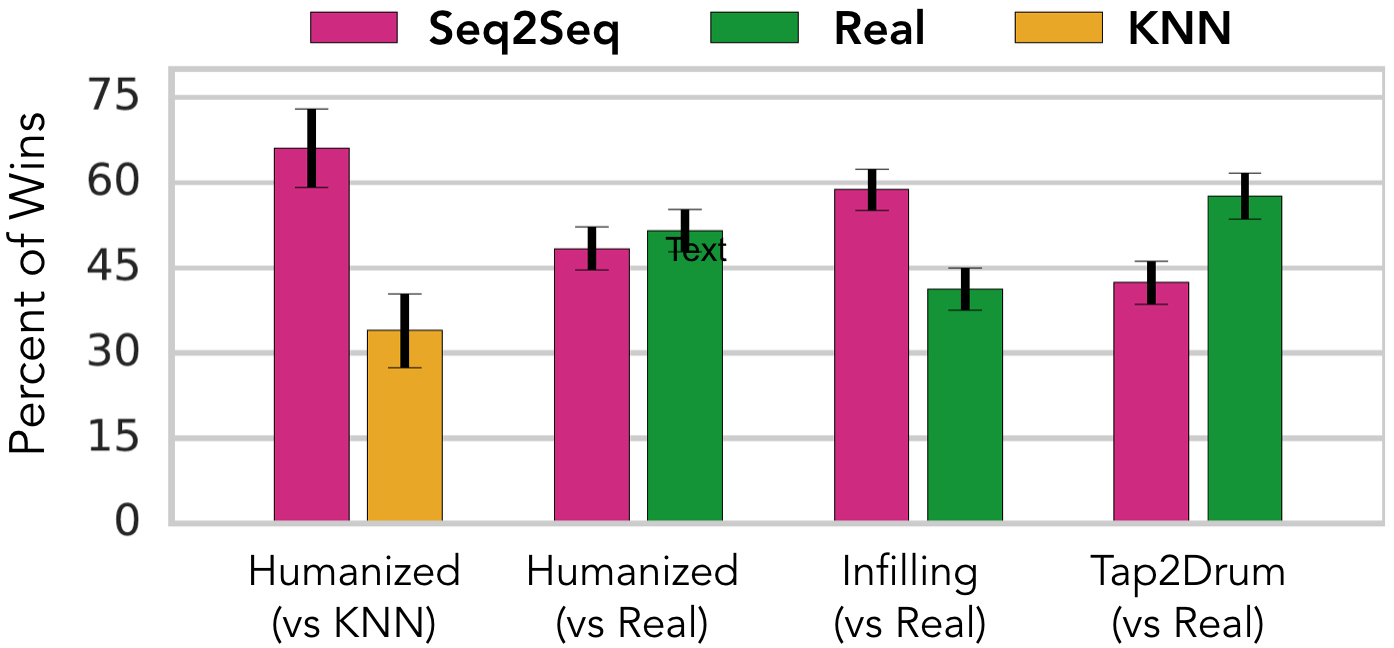}
    \caption{Results of head-to-head listening tests for different tasks and baselines, with 95\% confidence interval bands. The experiments included 56, 188, 189, and 177 comparisons, respectively.}
    \label{fig:listening_test}
    \end{centering}
\end{figure}

\paragraph{Humanization: Comparison with baseline.} For this experiment, we compare the Humanization model that we judged produced the best subjective outputs (Seq2Seq with VIB), with the best baseline model (KNN).  We randomly selected 32 2-measure sequences from the test set, removing all microtiming and velocity information, and then generated new performances of all 32 sequences using both Humanization models.  We presented participants with random pairs of clips, one of which was generated by each model, asking them to judge which clip sounds more like a human drummer.  The Seq2Seq model significantly outperformed the baseline as can be seen in the first column of Figure~\ref{fig:listening_test}.

\paragraph{Comparison with real sequences from the test set.}

Perhaps a more compelling test of the real-world viability of our models is to ask listeners to compare generated outputs with clips of real drum performances; if the model outputs are competitive, this suggests that the generated drums are perceptually comparable with real performances.  We structured this test in the same way as the baseline comparison, asking listeners which sequence sounds more like a human drummer; in this case each pair contains one real clip from the test set and one generated clip.  As noted in Section \ref{ref:ssec:preprocessing}, because our models do not generate drum rolls faster than 16th notes, we compared against the preprocessed versions of test set clips (which also do not have faster drum rolls) to ensure fair comparison.  Figure~\ref{fig:listening_test} summarizes the results of this test for each of our tasks (Humanization, Infilling, and Tap2Drum), showing the generated outputs from our Seq2Seq models are competitive with real data. 

%To get a sense of the upper bound on the realism and quality of our model outputs, we presented listeners with pairs of audio clips in which each pair contained one example generated by our model model and one real example drawn from the test set.  We asked participants to identify which example was played by a real drummer.  This test measures the degree to which listeners can distinguish between real and generated samples; if they are unable to tell the difference, this provides evidence that the model outputs contain the semantics and expressiveness of real performances.  We compared the generated outputs of our three models, Humanization, Infilling, and Tap2Drum against real data points from our test set; table \ref{real_vs_generated} summarizes the results of these surveys.

%We measure model performance both though quantitative metrics and through listening tests.  First, following previous work in expressive performance modeling, we measure the error between ground truth and predicted microtiming and dynamics.  We use several metrics for comparison, summarizing results for all models in Table \ref{humanization_results}.  As a baseline, we always predict $0$ for timing offsets and $0.5$ for velocity.

\subsection{Quantitative Metrics}

Though it is difficult to judge these generative models with simple quantitative metrics, we report several quantitative evaluations for comparison, summarizing results in Table \ref{humanization_results}, along with 95\% bootstrap confidence intervals.

%\begin{table*}
%\begin{center}
%\begin{tabular}{| l | c | c | c || c | c |} \hline
%Model & MAE (ms) & MSE (16th note) & Timing KL & Velocity KL  \\ \hline

%Baseline & 22.6 \small{[22.45--22.8]} & 0.041 \small{[0.041--0.042]}   & N/A  & N/A  \\ \hline
%Linear & 20.3 \small{[20.1--20.4]} & 0.035 \small{[0.034--0.036]} &  6.03 %\small{[5.89--6.17]} & 1.92 \small{[1.86--1.98]}  \\
%KNN & 20.9 \small{[20.7--21.1]}  & 0.037 \small{[0.037--0.038]} & 2.43 \small{[2.36--2.51]} & 0.233 \small{[0.216--0.249]} \\ \hline
%MLP & 19.6 \small{[19.5--19.8]}& 0.033 \small{[0.033--0.034]} & 7.91 \small{[7.64--8.14]} & 1.83 \small{[1.75--1.90]}  \\ 
%Seq2Seq & \textbf{18.7 \small{[18.6--18.9]}} & \textbf{0.030 \small{[0.029--0.030]}} & \textbf{1.64 \small{[1.59--1.71]}} & 0.519 \small{[0.489--0.550]} \\ 
%Seq2Seq + VIB & 19.3 \small{[19.1--19.5]} & 0.030 \small{[0.030--0.031]} & 1.99 \small{[1.90--2.06]} & 0.141 \small{[0.129--0.153]}  \\
%Groove Transfer & 22.05 \small{[20.5--23.6]} & 0.038 \small{[0.032--0.042]} & 1.70 \small{[1.64--1.76]} & \textbf{0.138 \small{[0.133--0.144]}} \\ \hline

%\end{tabular}
%\end{center}
%\caption{\label{humanization_results} Metrics for different Humanization models, with 95\% bootstrap confidence intervals.}
%\end{table*}

\begin{table*}
\begin{center}
\begin{tabular}{| l | c | c | c || c | c |} \hline
Model & MAE (ms) & MSE (16th note) & Timing KL & Velocity KL  \\ \hline

Baseline & 22.6 \small{[22.45--22.72]} & 0.041 \small{[0.041--0.042]}   & N/A  & N/A  \\ \hline
Linear & 19.77 \small{[19.63--19.88]} & 0.033 \small{[0.033--0.034]} &  4.79 \small{[4.68--4.88]} & 1.70 \small{[1.66--1.74]}  \\
KNN & 22.34 \small{[22.19--22.45]}  & 0.043 \small{[0.042--0.0438]} & 1.10 \small{[1.07--1.12]} & 0.53 \small{[0.51--0.56]} \\ \hline
MLP & 19.25 \small{[19.13--19.40]}& 0.032 \small{[0.031--0.032]} & 7.62 \small{[7.44--7.80]} & 2.22 \small{[2.16--2.29]}  \\ 
Seq2Seq & 18.80 \small{[18.67--18.90]} & 0.032 \small{[0.031--0.032]} & 0.31 \small{[0.31--0.33]} & \textbf{0.08 \small{[0.08--0.09]}} \\ 
Seq2Seq + VIB & \textbf{18.47 \small{[18.37--18.60]}} & \textbf{0.028 \small{[0.028--0.029]}} & 2.80 \small{[2.72--2.86]} & 0.22 \small{[0.21--0.23]}  \\ \hline
Groove Transfer & 25.04 \small{[24.82--25.28]} & 0.052 \small{[0.051--0.053]} & \textbf{0.24 \small{[0.23--0.25]}} & 0.12 \small{[0.12--0.13]} \\
Groove Transfer + VIB & 24.49 \small{[24.25--24.72]} & 0.051 \small{[0.049--0.052]} & 0.27 \small{[0.26--0.28]} & 0.20 \small{[0.19--0.20]} \\ \hline

\end{tabular}
\end{center}
\caption{\label{humanization_results} Metrics for different Humanization models, with 95\% bootstrap confidence intervals.}
\end{table*}

\paragraph{Timing MAE.}  We report mean absolute error in milliseconds, which is useful for interpreting results in the context of studies on Auditory Temporal Resolution, a measure of the minimum amount of time required for the human ear to perceive a change in sound.  Studies show that temporal resolution depends on the frequency, loudness, and envelope of the sound as well as on the listener and type of recognition test (e.g., noise or pitch recognition)~\cite{muchnik1985minimal, kumar2016temporal}.  On tests for which the ear is more sensitive, such as the Gap-in-Noise test, mean values can be as low as 2ms, while for pitched tests like Pure Tone Discrimination, values can be 20ms or more~\cite{an2014effects}.  Most likely, the resolution at which the ear can perceive differences in drum set microtiming lies somewhere in between.

\paragraph{Timing MSE.}  Following~\citet{wright2006towards}, for another perspective on timing offsets, we look at mean squared error relative to tempo, here using fractions of a 16th note as units.  Since beats are further apart at slower tempos, this metric weights errors equally across all tempos.

\paragraph{Velocity KL / Timing KL.}  One drawback of the above metrics, which are aggregated on a per-note basis, is that they do not account for the possibility of mode collapse or blurring when comparing methods~\cite{trieu2018improvising}.  The effects of blurring seem to be particularly severe for velocity metrics; instead of averaging velocity errors across all notes, previous work computes similarity between the distributions of real and generated data ~\cite{tidemann2007imitating, hellmer2015quantifying}.  We adopt this approach, first predicting velocities and offsets for the entire test set and then comparing these with ground truth distributions.  For these metrics, we aggregate all predicted notes into four groups based on which 16th note position they align with.  We calculate the means and standard deviations for each group of notes, compute the KL Divergence between predicted and ground truth distributions based on those means and standard deviations, and then take the average KL Divergence across the four groups.  These distribution based metrics should not be treated as a gold standard either, but they do tend to penalize severe instances of blurring or mode collapse, as can be seen with the Linear and MLP models.

\section{Analysis}

\subsection{Comparisons with KNN baseline}

Based on the results of the listening tests shown in Figure~\ref{fig:listening_test}, Seq2Seq models clearly offer a powerful method for generating expressive drum performances.  The listener preference for Humanization using Seq2Seq over KNN is substantial, and moreover, these survey participants were not specifically chosen from a pool of expert musicians or drummers; that this pool of listeners was able to so clearly identify the Seq2Seq models as more realistic than the baseline seems to indicate that the model captures important nuances that make drumming realistic and expressive.

\subsection{Comparing Humanization to real data}

The survey results indicate that, at least for our population of listeners, drum performances generated through Seq2Seq Humanization are difficult to distinguish from real data; statistically, the results show no significant difference.

\subsection{Comparing Infilling to real data}

Perhaps counter-intuitively, a significant fraction of listeners in this experiment (nearly 60\%) identified the generated outputs as sounding \emph{more} human than the real data.  One potential explanation for this result is that among our test data, some sequences sound subjectively better than others.  A small fraction of the recordings are from amateur drummers, who sometimes make mistakes or play at a lower level.  In replacing the original hi-hat parts, the Infilling model in effect resamples from the data distribution and may generate better sounding, more likely parts.  This result suggests a potential use for the model as a corrective tool that works by resampling parts of an input that have noise or imperfections.

\subsection{Comparing Tap2Drum to real data}

Figure~\ref{fig:listening_test} demonstrates the slight preference of listeners for the real data over performances generated by Tap2Drum (about 56\%).  This difference is significant but comparatively small relative to the difference between Seq2Seq and KNN Humanization, indicating that Tap2Drum may be a viable way of controlling expressive performances in practice.  More work is needed to better understand how much control this model offers and how people interact with the model in different contexts; qualitative research with musicians and music producers offers one path forward.

\subsection{Groove Transfer}

Evaluating Groove Transfer is challenging in the absence of existing methods for comparison; nonetheless, we believe that this particular version of style transfer yields subjectively interesting outputs and merits further investigation both in terms of its architecture and its potential for creative application in the future.

\subsection{Quantitative Results}

As might have been expected, the Seq2Seq models achieve the best results on the timing MAE and MSE metrics, while also outperforming the baselines on the distribution-based metrics.  The Groove Transfer models, in exchange for the added control given by the ability to perform a beat in the style of any other beat, sacrifice some accuracy on the Humanization task compared to Seq2Seq, as can be seen by the higher MAE error. 
%Though as with many generative models, simple quantitative metrics do not always align with subjective perception, we find that measuring the similarities between distributions of real and generated sequences proves to be a useful evaluation; to this end, we report on our Velocity KL and Timing KL metrics.

\section{Conclusions}

In this work, we demonstrate that learning inverse sequence transformations can be a powerful tool for creative manipulation of sequences. We present the Groove MIDI Dataset with an order of magnitude more expressive drum performances than comparable research datasets, new methods for generating expressive drum performances, and both quantitative and qualitative results demonstrating state-of-the-art performance on Humanization.

We also explore new applications, such as Tap2Drum, which may enable novices to easily generate detailed drum performances.  Our results raise the possibility of learning other creative inverse transformations for sequential data such as text and audio.  We hope this line of research will ultimately lead a variety of interesting creative applications, just as similar GAN-based techniques have done for images and video.

\section{Acknowledgements} 

Thank you to our drummers for their hard work and talent, without whom this work would not have been possible: Dillon Vado, Jonathan Fishman, Michaelle Goerlitz, Nick Woodbury, Randy Schwartz, Jon Gillick, Mikey Steczo, Sam Berman, and Sam Hancock.

Thank you to the anonymous reviewers and to Anna Huang, Chris Donahue, Colin Raffel, Curtis ``Fjord" Hawthorne, David Ha, Ian Simon, Mike Tyka, Monica Dinculescu, and Yingtao Tian for valuable advice and feedback.

%We hope that this line of research will benefit a variety of music creators. 
% In practice, many styles of contemporary recorded music are often made with a mix of physical and digital tools~\cite{askeroi2013reading}; music composers and producers, whether professional or amateur, write for ensembles that include instruments they themselves cannot play or do not have the means to record.  Instead, they sketch these parts in MIDI (Musical Instrument Digital Interface) format, generating the sounds of the instruments with synthesizers or samplers.  These generated parts, however, often lack the dynamics and expressiveness that make for compelling music.

\bibliography{groovae}
\bibliographystyle{icml2019}

\appendix

\section{GMD: Groove MIDI Dataset} \label{ref:sec:gmd}

The Groove MIDI Dataset (GMD), has several attributes that distinguish it from other data:

\begin{figure}
    \begin{centering}
    \includegraphics[scale=0.248]{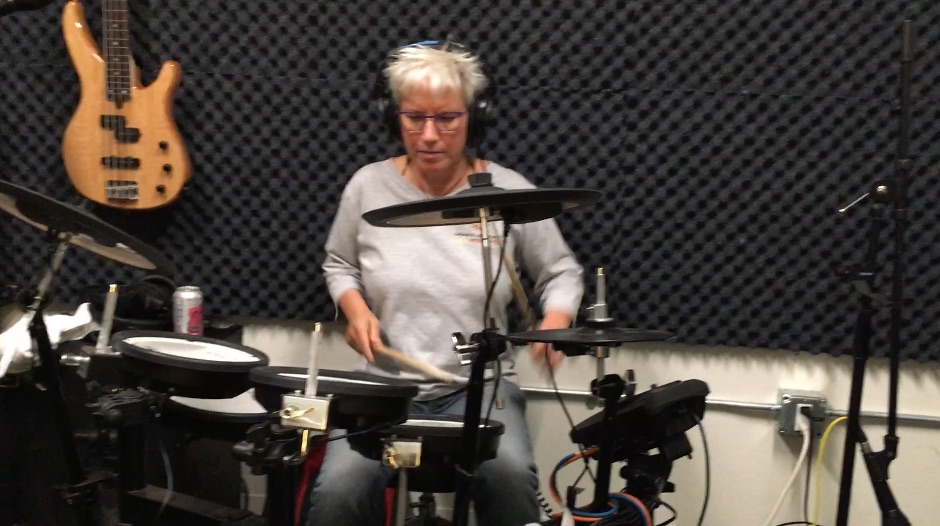}
    \caption{A drummer recording for the Groove MIDI Dataset}
    \label{fig:drummer}
    \end{centering}
\end{figure}

\begin{itemize}
    \item The dataset contains about 13.6 hours, 1,150 MIDI files, and 22,000 measures of drumming.

    \item Each performance was played along with a metronome set at a specific tempo by the drummer.  Since the metronome provides a standard measurement of where the musical beats and subdivisions lie in time, we can deterministically quantize all notes to the nearest musical division, yielding a musical score.  Recording to a metronome also allows us to take advantage of the prior structure of music by modeling relative note times (quarter note, eighth note, etc.) so as to free models from the burden of learning the concept of tempo from scratch.  The main drawback of the metronome is that we enforce a consistent tempo within each individual performance (though not across performances) so we do not capture the way in which drummers might naturally change tempo as they play.
    
    \item The data includes performances by a total of 10 drummers, 5 professionals and 5 amateurs, with more than 80\% coming from hired professionals.  The professionals were able to improvise in a wide range of styles, resulting in a diverse dataset.
    
    \item The drummers were instructed to play a mix of long sequences (several minutes of continuous playing) and short beats and fills. 
    
    \item Each performance is annotated with a genre (provided by the drummer), tempo, and anonymized drummer ID.
    
    \item Most of the performances are in 4/4 time, with a few examples from other time signatures; we use only the files in 4/4 in this work.
    
    \item In addition to the MIDI recordings that are the primary source of data for the experiments in this work, we captured the synthesized audio outputs of the drum set and aligned them to within 2ms of the corresponding MIDI files.  These aligned audio files may serve as a useful resource for future research in areas like Automatic Drum Transcription.
    
    \item A train/validation/test split configuration is provided for easier comparison of model accuracy on various tasks.   
    
    \item Four drummers were asked to record the same set of 10 beats in their own style.  These are included in the test set split, labeled `eval-session/groove1-10'.
    
\end{itemize}

\begin{table*}
\begin{center}
\begin{tabular}{| l | c | c | c | c |} \hline
Pitch & Roland Mapping & GM Mapping & Drum Category & Count \\ \hline
36 & Kick & Bass Drum 1 & Bass (36) & 88067 \\
38 & Snare (Head) & Acoustic Snare & Snare (38) & 102787 \\
40 & Snare (Rim) & Electric Snare & Snare (38) & 22262 \\
37 & Snare X-Stick & Side Stick & Snare (38) & 9696 \\
48 & Tom 1 & Hi-Mid Tom & High Tom (50) & 13145 \\
50 & Tom 1 (Rim) & High Tom & High Tom (50) & 1561 \\
45 & Tom 2 & Low Tom & Low-Mid Tom (47) & 3935 \\
47 & Tom 2 (Rim) & Low-Mid Tom & Low-Mid Tom (47) & 1322 \\
43 & Tom 3 (Head) & High Floor Tom & High Floor Tom (43) & 11260 \\
58 & Tom 3 (Rim) & Vibraslap & High Floor Tom (43) & 1003 \\
46 & HH Open (Bow) & Open Hi-Hat & Open Hi-Hat (46)	& 3905 \\
26 & HH Open (Edge) & N/A & Open Hi-Hat (46) & 10243 \\
42 & HH Closed (Bow) & Closed Hi-Hat & Closed Hi-Hat (42) & 31691 \\
22 & HH Closed (Edge) & N/A & Closed Hi-Hat (42) & 34764 \\
44 & HH Pedal & Pedal Hi-Hat & Closed Hi-Hat (42) & 52343 \\
49 & Crash 1 (Bow) & Crash Cymbal 1 & Crash Cymbal (49)	& 720 \\
55 & Crash 1 (Edge) & Splash Cymbal & Crash Cymbal (49) & 5567 \\
57 & Crash 2 (Bow) & Crash Cymbal 2 & Crash Cymbal (49) & 1832 \\
52 & Crash 2 (Edge) & Chinese Cymbal & Crash Cymbal (49) & 1046 \\
51 & Ride (Bow) & Ride Cymbal 1 & Ride Cymbal (51) & 43847 \\
59 & Ride (Edge) & Ride Cymbal 2 & Ride Cymbal (51) & 2220 \\
53 & Ride (Bell) & Ride Bell & Ride Cymbal (51) & 5567 \\

\hline
\end{tabular}
\caption{\label{tab:drum_categories} List of Drum Categories}
\end{center}
\end{table*}

\section{Reduced MIDI Mapping} \label{ref:sec:midimap}

The Roland TD-11 used to record the performances in MIDI uses some pitch values that differ from the General MIDI (GM) Specifications.  Table~\ref{tab:drum_categories} displays the choice of MIDI notes to represent the nine essential drum voices for this study, along with the counts of each pitch in the data.

\end{document}